\documentclass[aps,twocolumn,prd]{revtex4}
\usepackage{graphicx}

\begin{document}

\title{Born-Infeld corrections to Coulombian interactions}

\author{Rafael Ferraro}
\email{ferraro@iafe.uba.ar}
\thanks{Member of Carrera del Investigador Cient\'{\i}fico (CONICET,
Argentina)} \affiliation{Instituto de  Astronom\'\i a y F\'\i sica
del Espacio, Casilla de Correo 67, Sucursal 28, 1428 Buenos Aires,
Argentina} \affiliation{Departamento de F\'\i sica, Facultad de
Ciencias Exactas y Naturales, Universidad de Buenos Aires, Ciudad
Universitaria, Pabell\'on I, 1428 Buenos Aires, Argentina}

\author{Mar\'{\i}a Evangelina Lipchak}
\email{mariaevangel@gmail.com} \affiliation{Departamento de F\'\i
sica, Facultad de Ciencias Exactas y Naturales, Universidad de
Buenos Aires, Ciudad Universitaria, Pabell\'on I, 1428 Buenos
Aires, Argentina}

\begin{abstract}
Two-dimensional Born-Infeld electrostatic fields behaving as the
superposition of two point-like charges in the linearized
(Maxwellian) limit are worked out by means of a non-holomorphic
mapping of the complex plane. The changes underwent by the
Coulombian interaction between two charges in Born-Infeld theory
are computed. Remarkably, the force between equal charges goes to
zero as they approach each other.
\end{abstract}



\maketitle

\section{Introduction}

When forces between {\it charges} are considered in nonlinear
theories, the picture of the field due to a charge acting on
another charge is no longer applicable. Since the superposition
principle is not feasible, a multiple charge configuration has to
be analyzed as a new problem instead of the mere addition of
already known solutions. Even the expression ``multiple charge
configuration" calls for an explanation. For those theories
behaving as a linear theory in the regime of weak field, a
multiple charge configuration can be defined as a solution of the
nonlinear field equations going to a superposition of individual
(linear) charges at infinity. Given a static multiple charge
solution of some nonlinear theory, the force on a charge can be
worked out by computing the flux of the stress tensor through a
surface surrounding the charge. Since the stress tensor is
divergenceless, the force will result different from zero only at
those points where the stress tensor is singular, which provides a
way of localizing the charges.

This article is aimed to solve two-dimensional static
configurations of two charges in Born-Infeld nonlinear
electrodynamics, and compute the interaction strength. In Section
II we summarize the Born-Infeld theory. In Section III we
characterize the two-dimensional electrostatic solutions by means
of a non-holomorphic complex transformation. In Section IV we
obtain the repulsive and attractive interaction between equal and
opposite charges. We compute corrections to the Coulombian
interaction for distant charges, and show that the repulsive force
vanishes when equal charges approach each other. The conclusions
are displayed in Section V.

\section{Born-Infeld electrodynamics}
Born-Infeld electrodynamics is a non-linear theory whose initial
objective was to render finite the self-energy of a point-like
charge. In Born-Infeld electrostatics the electric field {\bf E}
due to a point-like charge does not diverge but goes to a finite
value $b$ at the charge position. The energy-momentum tensor still
diverges at the charge position but the the integration of the
energy density becomes finite. The fundamental constant $b$ is an
upper bound for the fields, and regulates the transition to the
weak field regime: for fields much smaller than $b$ the theory
behaves as Maxwell electromagnetism. By healing the field of
singularities, Born and Infeld pretended that the theory could be
regarded from a unitary standpoint: the only physical entity would
be the field, whereas the charges would be just a part of the
field \cite{1,2,3,4,5}. They even believed that the solutions will
contain some essential features of the charge dynamics, which is
by no means true since the theory allows for static multiple
charge solutions.

Born-Infeld electrodynamics possesses outstanding physical
properties: together with Maxwell theory, they are the only spin 1
field theories having causal propagation \cite{6,7} and absence of
birefringence \cite{6,8}. Although concrete solutions for
propagating Born-Infeld electromagnetic waves are not sufficiently
known --apart from trivial free waves solutions--, solutions for
waves propagating in static background fields and waveguides have
been recently obtained \cite{9,10}. The renewed interest in
Born-Infeld theory can be traced to its emergence in the study of
strings and branes: loop calculations for open superstrings lead
to a Born-Infeld type low energy action \cite{11,12,13}. Nowadays
Born-Infeld-like Lagrangians have been proposed for quintessential
matter models and inflation \cite{14}, and also for alternative
theories of gravity \cite{15}. Born-Infeld charges coupled with
gravity have been investigated in attempts to remove geometrical
singularities of charged black holes \cite{16}.

Born-Infeld Lagrangian density for the electromagnetic field
$F_{ij} =\partial _i A_j -\partial _j A_i$ is \cite{2}
\begin{eqnarray}
\nonumber L_{BI} =-\frac{1}{4\pi \,c}\left( {\sqrt {\left| {\det
\,(b\,g_{i{\kern 1pt}j} +F_{i{\kern 1pt}j} )} \right|} -\sqrt
{\left| {\det \,(b\,g_{i{\kern 1pt}j} )} \right|} } \right)\\
=\;\,\frac{\sqrt {-g} }{4\pi \,c}\,\,b^2\,\left( {1-\sqrt
{\,1+b^{-2}2S-b^{-4}P^2} } \right)\label{eq3}
\end{eqnarray}
where $S$ and $P$ are the scalar and pseudo-scalar invariants,
\begin{eqnarray}
\nonumber &S=\frac{1}{4}\,F_{ij} \,F^{ij}=\frac{1}{2}\,\left(
B^2-E^2 \right) \\ &P=\frac{1}{4}\,^\ast F_{ij} \,F^{ij}={\rm {\bf
E}}\cdot {\rm {\bf B}}\ , \label{eq4}
\end{eqnarray}
and $b$ is a new universal constant with units of field, which
plays the role of upper bound for the electrostatic field. The
term $\sqrt {\left| {\det \left( {b\,g_{ij} } \right)} \right|}$
in (\ref{eq3}) makes $L_{BI}$ to vanish when the electromagnetic
field vanishes. $L_{BI}$ goes to Maxwell Lagrangian density in the
limit $b \to \infty$. By defining the 2-form
\begin{equation}
\label{eq5} {\mathcal F}_{i\, j} = \frac{F_{i\, j} -b^{-2}
P\,\,^\ast F_{i\, j} }{\sqrt{1 + b^{-2} 2\, S - b^{-4} P^2}} ,
\end{equation}
the Euler-Lagrange equations coming from $L_{BI}$ are written as
\begin{equation}
\label{eq1} d\,^\ast {\mathcal F} = 0
\end{equation}
These equations are supplemented with the identities $dF = 0$
(i.e., $F_{[j k, i]} =0$), since the field is an exact 2-form ($F
= dA$). The energy-momentum tensor is
\begin{equation}
T_{ij} =-\frac{1}{4\pi }\,F_{ik}{\mathcal F}_j ^{\
k}-\frac{b^2}{4\pi } g_{ij}\left( {1-\sqrt
{1+b^{-2}2\,S-b^{-4}P^2} } \right) \label{eq6}
\end{equation}
which verifies the energy-momentum conservation,
\begin{equation}
T^l_{\, \, k\, ;\, l} =0,\label{eq7}
\end{equation}
at all the places where $T_{ij}$ is non-singular. For a point-like
charge $Q$, ${\mathcal F}= Q\, r^{-2}\, dt\wedge dr$ diverges at
the charge position but $F$ is finite. Thus, although the energy
density still diverges at the charge position, the integrated
energy turns out to be finite.

Owing to the non-linear character of the field equations
(\ref{eq5}), it is hard to find explicit solutions in an analytic
way, apart from highly symmetric configurations. In more general
cases, the solutions are displayed only in an implicit form
\cite{17,18}.

\section{Two-dimensional electrostatic solutions}
Two-dimensional electrostatic solutions, together with conditions
guaranteeing uniqueness of solutions, have been worked out in a
rather cumbersome parametric way by resorting to the relationship
between minimal surface equations and the Born-Infeld
electrostatic problem \cite{19,20}. Recently, two-dimensional
electrostatic solutions in Euclidean space have been obtained by
using a non-holomorphic transformation of the complex plane. This
method has been used for working out the field lines and
self-energies of point-like two-dimensional multipoles \cite{21}.
The method searches a coordinate transformation in the plane, $(x,
y)\rightarrow (u, v)$, such that $(u, v)$ are orthogonal
coordinates and $u(x, y)$ results to be the potential for a
Born-Infeld electrostatic field: ${\bf E}=-\nabla u$ (then the
coordinate lines are equipotential and field lines respectively).
The equation to be fulfilled by $u(x, y)$ is the one resulting
from Eq.(\ref{eq1}) when the field $F$ is chosen to be the exact
2-form $F=du\wedge dt$, corresponding to the electrostatic
potential $A = u(x,y) dt$. In current language, the equation is
\begin{equation}
\label{eq11} \nabla .\,\left( {\frac{\nabla \,u(x,y)}{\sqrt
{1-b^{-2}\,\left| {\nabla \,u(x,y)} \right|^2} }} \right)=0
\end{equation}
which becomes the Laplace equation when $b\to\infty $. Actually
this equation can be also obtained by directly suppressing the
time coordinate $t$ and the third spatial coordinate in the
Lagrangian. Thus, in two dimensions the (static) field is the
exact 1-form $F = du$. In addition, $^*{\mathcal F}$ in
Eq.(\ref{eq1}) is the 1-form $^*{\mathcal
F}={^*F}/\sqrt{1-b^{-2}E^2}$.

In terms of complex numbers $z = x + i\, y$ and $w = u + i\, v$,
the coordinate transformation $(x, y)\rightarrow (u, v)$ can be
regarded as $z\rightarrow z(w,\overline w)$ or
\begin{equation}
dz = p(w,\overline w)\, dw + q(w,\overline w)\, d\overline
w\label{e1}
\end{equation}
(the bar means complex conjugate). The integrability condition for
Eq.(\ref{e1}) requires that
\begin{equation}
{\overline\partial}p(w,\overline w)=\partial q(w,\overline w)\
,\label{ee1}
\end{equation}
where $\partial$ and $\overline\partial$ are exterior derivatives
with respect to $w$ and $\overline w$ (Dolbeault operators). To
glance the Euclidean metric in $(u, v)$ coordinates, one can write
\begin{equation}
dx^2+dy^2=dz\, d\overline z=(|p|^2+|q|^2)\, |dw|^2+2\,
Re(p\,\overline q\, dw^2) \label{e3}
\end{equation}
To get orthogonal coordinates $(u, v)$, the terms containing $du\,
dv$ must be ruled out from the quadratic form (\ref{e3}). Then
\begin{equation}
Im(p\, \overline q)=0\ .\label{ee2}
\end{equation}
The 1-form $dw$ can be solved from Eq.(\ref{e1}) and its complex
conjugate:
\begin{equation}
dw=\frac{\overline p\, dz\, -\, q\, d\overline
z}{|p|^2-|q|^2}\label{e4}
\end{equation}
In two-dimensional Euclidean space it is $^*dz=i\, dz$; then
\begin{equation}
^*dw=i\,\frac{\overline p\, dz\, +\, q\, d\overline
z}{|p|^2-|q|^2}\label{e5}
\end{equation}
In order that the field $^*{\mathcal
F}=Re(^*dw)/\sqrt{1-b^{-2}E^2}$ accomplishes the Eq.(\ref{eq1}),
we have to properly choose the functions $p$ and $q$. Let us try
the choice $q=0$. Then, the integrability condition (\ref{ee1})
implies that $p=p(w)$ ($w$ is a holomorphic function of $z$). In
this case $^*dw = i\, dw$ is an exact 1-form. So, Eq.(\ref{eq1})
is satisfied only if $b\to\infty$ (because it will result
$^*{\mathcal F} = {^*F} = Re(^*dw)$). This is the case of the
Coulombian field (it is well known that holomorphic functions
provide solutions for Laplace equation in two dimensions). The
Coulombian solution is
\begin{equation}
\label{e6} {\bf E}=E_x +i\,E_y =-\frac{\partial u}{\partial
x}-i\,\frac{\partial u}{\partial y}=-\frac{1}{\ \overline{p(w)}\
}.
\end{equation}

Let us try a choice of $p$ and $q$ leading to $q=0$ when
$b\to\infty$. We guess the choice \cite{21}
\begin{equation}
p(w)\, \overline{q(\overline w)}\ =\ \frac{1}{4\, b^2}\label{e7}
\end{equation}
(notice that the integrability and orthonormality conditions
(\ref{ee1}, \ref{ee2}) are satisfied). Then
\begin{equation}
du = Re(w) = 4\,b^2\, \frac{Re(p)\, dx\, +\, Im(p)\, dy}{4\,
b^2\,|p|^2\,+\,1}\label{e8}
\end{equation}
\begin{equation}
dv = Im(w) = 4\,b^2\,\frac{-Im(p)\, dx\, +\, Re(p)\, dy}{4
\,b^2\,|p|^2\,-\,1}\label{e9}
\end{equation}
Therefore
\begin{equation}
\label{eq14} {\bf E}=E_x +i\,E_y =-\frac{\partial u}{\partial
x}-i\,\frac{\partial u}{\partial
y}=-\frac{4\,b^2\,p(w)}{4\,b^2\,|p(w)|^2+1}.
\end{equation}
and
\begin{equation}
\label{e11} \sqrt{1-\frac{E^2}{b^2}}\, =\,
\pm\,\frac{\,4\,b^2\,|p(w)|^2\,-\,1\,}{\,4\,b^2\,|p(w)|^2\,+\,1\,}.
\end{equation}
These results imply that $^*{\mathcal F}$ is an exact 1-form. In
fact
\begin{equation}
^*{\mathcal F}\,=\,\frac{^*du}{\sqrt{1-b^{-2}E^2}}\,=\,\pm\, i\,
dv\label{e10}
\end{equation}
Therefore, Eq.(\ref{eq1}) is satisfied.

Eq.(\ref{eq14}) shows the way to generate those Born-Infeld field
behaving in the weak field limit (far from the charges) like a
given Coulombian field: choose the holomorphic function $p(w)$
associated to the Coulombian field (\ref{e6}) and replace it in
(\ref{eq14}). However, as the field should be expressed as a
function of $z$ or $(x,y)$ instead of $w$ or $(u, v)$, we have to
take into account that the relation between $z$ and $w$ is no
longer the Coulombian relation; according to Eq.(\ref{e1}) the
relation now is
\begin{equation}
dz = p(w)\, dw + \frac{1}{4\, b^2\, \overline{p(w)}\ }\,
d\overline w\label{e12}
\end{equation}
This equation amounts to a non-holomorphic relation between $z$
and $w$. The electrostatic potential fulfilling the Born-Infeld
equation (\ref{eq11}) is $u(x, y) = Re(w(z,\overline z))$.
Function $p(w)$ in Eq.(\ref{e12}) plays the role of a Coulombian
seed to obtain the Born-Infeld potential.

Eq.(\ref{eq14}) shows that $\vert{\bf E}\vert$ does not diverge
but attains its maximum value $b$ at the points where
\begin{equation}
\label{eq15} |p(w)|\ =\ \frac{1}{2\,b}
\end{equation}
According to Eq.(\ref{e7}), at these points it is $|p|=|q|$; so
they are singular points of the coordinate change (\ref{e4},
\ref{e5}). The relation (\ref{eq15}) describes the curve where the
energy-momentum tensor (\ref{eq6}) is singular, because the
vanishing of (\ref{e11}) implies that ${\mathcal F}$ diverges.
This curve is then the charge location. If the curve
$|p(w)|=(2b)^{-1}$ is closed, then it separates two different
regions in the complex plane: i) $|p(w)|>(2b)^{-1}$ and ii)
$|p(w)|<(2b)^{-1}$. Only the first region can realize the
Coulombian limit $E/b \to 0$. Since the Born-Infeld field
(\ref{eq14}) should go to the Coulombian field (\ref{e6}) at
infinity, the region $|p(w)|>(2b)^{-1}$ corresponds to the
exterior of the charge distribution. The curves where
$|p(w)|=(2b)^{-1}$ have been studied in Ref. \cite{21} for the
configurations associated with Coulombian multipoles. In these
cases the curves are closed and turn out to be epicycloids whose
sizes are determined by $b$ and the multipolar moment. One can say
that Born-Infeld field smoothes singularities in two different
ways: on one hand it smoothes the divergence of the
energy-momentum tensor in order that the self-energy results to be
finite; on the other hand the point-like character of the
Coulombian multipoles is spread to the surface (in this case a
curve) where the field reaches the upper bound $b$.

\section{Force between two-dimensional monopoles}
We are going to study the electrostatic Born-Infeld configuration
corresponding to two point-like monopoles separated by a distance
$d$ in the Coulombian limit. The interaction force between charges
will result from the momentum flux through a closed surface $S$
containing one of the charges. We will choose the $x$-axis along
the line joining the charges, and the origin of coordinates at the
intermediate point. Since the symmetry dictates that the force is
directed along the $x$-axis, then the involved momentum flux is
$T^{ x j }$\textit{dS}$_{ j}$. As usual, we choose $S$ as the
surface formed by the $y$-axis, and a semi-circumference of
infinite radius centered at the origin of coordinates; on this
last surface the flux is null. Thus
\begin{eqnarray}
\label{eq16} &&F^x=-\int\limits_{y-axis} T^{xx}\;n_x \;dS\\
\nonumber &&=\frac{b^2}{4\pi }\int\limits_{-\infty }^{+\infty }
\left[{1-\left(1-\frac{E^2}{b^2}\right)^{\frac{1}{2}} -\frac{E_x
^2}{b^2}\left( {1-\frac{E^2}{b^2}} \right)^{-\frac{1}{2}}} \right]
_{x=0} dy.
\end{eqnarray}
For repulsive interactions between equal charges it is $E_{x} = 0$
($v$ = \textit{constant}) and $dy = (\partial y/\partial u)\, du$,
at $x = 0$.  Thus, according to Eqs. (\ref{eq14}, \ref{e11},
\ref{e12}) the force is
\begin{equation}
 F^x=\frac{1}{8\pi}\int\limits_{y-axis}^{} {\frac{Im\left(p(w)
\right)}{\left|p(w)\right|^2}\ du}\label{eq17}
\end{equation}
For attractive interactions between opposite charges it is
$E_{x}=\pm E$ ($u$ = \textit{constant}) and $dy = (\partial
y/\partial v)\, dv$, at $x = 0$. Then the force is
\begin{equation}
\label{eq18} F^x=\frac{1}{8\pi}\int\limits_{y-axis}^{}
{\frac{Re\left(p(w)\right)}{\left|p(w)\right|^2} \ dv} .
\end{equation}

\subsection{Equal charges}
\begin{figure}[ht]
\centering
\includegraphics[scale=.4]{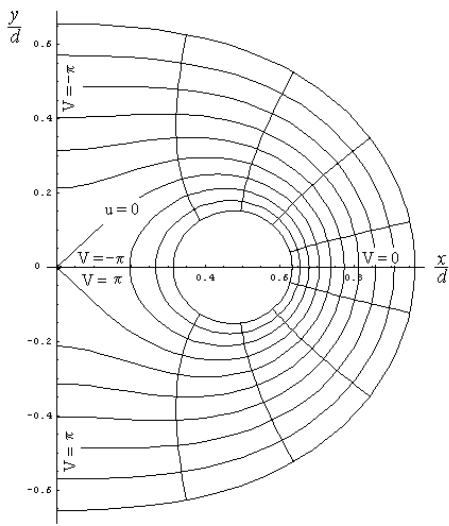}
\caption[]{Equipotencial and field lines for the Coulombian equal
charges (right semi-space). $V$ stands for $v/(2\,
\lambda)$.}\label{Fig1}
\end{figure}

\begin{figure}[ht]
\centering
\includegraphics[scale=.4]{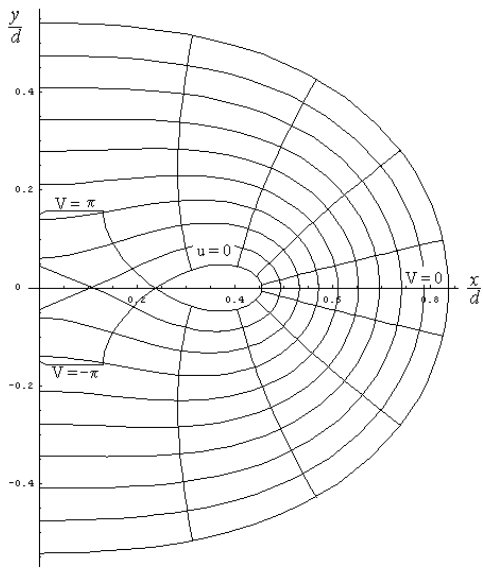}
\caption[]{Multi-valued Born-Infeld mapping (\ref{eq21}) for
$-\pi<v/(2\, \lambda)<\pi$ ($\alpha^2=40$). In the figure $V$
stands for $v/(2\, \lambda)$.}\label{Fig2}
\end{figure}

\begin{figure}[ht]
\centering
\includegraphics[scale=.8]{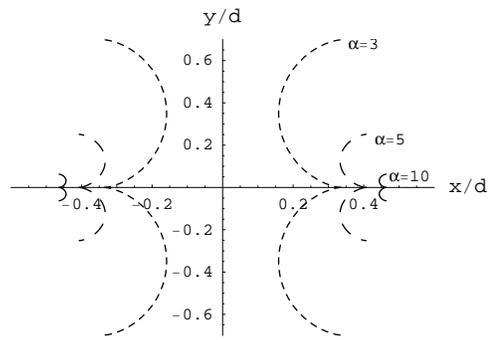}
\caption[]{Curves $|p(w)|=(2b)^{-1}$ characterized by
$\alpha=bd/\lambda$ for the configuration of two equal charges.
Only the cusps, but not the lobes, belong to the branch of $w(z)$
under consideration.}\label{Fig3}
\end{figure}

The well known Coulombian potential $u(x,y)$ for the repulsive
configuration of two equal charges $\lambda $ at a distance $d$
can be written as $u(x,y) = \it{Re}[w(z)]$, where $w(z)$ is the
holomorphic function
\begin{equation}
\label{eq19} w=-2\lambda \,\ln \left[ {\left( {\frac{2z}{d}}
\right)^2-1} \right].
\end{equation}
In Eq.(\ref{eq19}) a proper choice of the integration constant was
done in order that the Coulombian potential be zero at the origin.
Relation (\ref{eq19}) can be inverted to obtain the holomorphic
function $p$ characterizing the Coulombian mapping:
\begin{equation}
\label{eq20} z =\pm \frac{d}{2}\,\sqrt {\exp \left(
{-\frac{w}{2\lambda }} \right)+1} \quad ,
\end{equation}
where $\pm$ alludes to $x>0$ and $x<0$ regions. By differentiating
the Eq.(\ref{eq20}) one obtains
\begin{equation}
\label{e13}p(w)=\mp\frac{d}{8\, \lambda}\,\frac{\exp\left(
{-\frac{w}{2\lambda }} \right)}{\sqrt{\exp\left(
{-\frac{w}{2\lambda }} \right)+1}}
\end{equation}
By substituting in Eq.(\ref{e12}) and then integrating it, one
obtains the mapping leading to the Born-Infeld complex potential
$w(z, \overline z)$:
\begin{eqnarray}
\nonumber &&z=\pm \frac{d}{2}\,\Bigg[\sqrt{\exp\left(
-\frac{w}{2\lambda} \right)+1} \\ \nonumber &&-\frac{8}{\alpha
^2}\,\Bigg(\exp\left(\frac{\overline w}{2\lambda }
\right)\,\,\sqrt {\exp\left(-\frac{\overline
w}{2\lambda}\right)+1}
\\ &&+\frac{\overline w}{4\lambda}+\ln\left( 1+\sqrt{\exp\left(
-\frac{\overline w}{2\lambda}\right)+1}\right) \Bigg)
\Bigg]\label{eq21}
\end{eqnarray}
where $\alpha$ is the non-dimensional parameter $\alpha = b d/
\lambda$. In Eq.(\ref{eq21}) one recognizes the holomorphic
(Coulombian) seed (\ref{eq20}) and the anti-holomorphic
Born-Infeld correction. Differring from the Coulombian case
(\ref{eq20}), Eq.(\ref{eq21}) does not provide a unique
electrostatic potential $u$ to each point of the complex plane. In
particular, while the Coulombian mapping (\ref{eq20}) is periodic
in the $v$ coordinate, the Born-Infeld mapping (\ref{eq21}) fails
to be periodic because of the presence of the linear term
$\overline w/(4\lambda)$. Figure \ref{Fig1} shows the $u, v$ lines
for the Coulombian potential (\ref{eq20}). The equipotential $u=0$
passes through the coordinate origin. The lines
$v/(2\lambda)=\pm\pi$ coincides with the $y$-axis ($u>0$) and the
part of the $x$-axis joining the charges ($u<0$). The line $v=0$
is the piece of the $x$-axis going from the charges to infinite.
Figure \ref{Fig2} shows the $u, v$ lines resulting from the
Born-Infeld mapping (\ref{eq21}) for $-\pi<v/(2\lambda)<\pi$.
Still it is $v/(2\lambda)=\pm\pi$ at the $y$-axis; however, as a
consequence of the lost of the periodicity, the lines
$v/(2\lambda)=\pm\pi$ do not end at the $x$-axis but cross it, so
giving rise to a multi-valued figure for the potential $u(x, y)$.
To have a single-valued potential, the $u-v$ domain of the mapping
(\ref{eq21}) should be restricted by cutting it at the $x$-axis
(owing to this branch cut, the field is not continuous along the
line joining the charges). Like the Coulombian case, the line
$v=0$ is the piece of the $x$-axis from the charges to infinite.
Function $p(w)$ is real on the line $v=0$ (see Eq.(\ref{e13})),
and attains its maximum value $1/(2\, b)$ at the charges, where
the potential results to be
\begin{equation}
\exp\left(-\frac{u_{max}}{2\, \lambda}\right)=\frac{4}{\alpha^2}\,
\left(2+\sqrt{4+\alpha^2}\right)\label{umax}
\end{equation}
Since the equipotential lines surround the charges, the value
(\ref{umax}) is the maximum value for the Born-Infeld potential.
Notice that, by solving the equation $|p(w)|=1/(2\, b)$ for any
value of $v$ one obtains the curve $u(v)$ described by the
relation
\begin{eqnarray}
\label{eq22} &&\left( {\frac{\alpha }{4}} \right)^4\exp \left(
-{\frac{2u}{\lambda }} \right)-\exp \left( -{\frac{u}{\lambda }}
\right)\\ \nonumber &&\ \ \ \ \ \ \ \ -2\cos \left(
{\frac{v}{2\lambda }} \right)\exp \left( -{\frac{u}{2\lambda }}
\right)=1.
\end{eqnarray}
This curve decomposes in two parts at each side of the $y$-axis,
displaying cusps on the $x$-axis at $\vert x \vert < d/2$ (see
Figure \ref{Fig3}). In spite of the appearances, the charge
configuration is not spread on the lobes of Figure \ref{Fig3}, but
concentrates at the cusps. In fact, Eq.(\ref{eq22}) can only be
fulfilled for $u\geq u_{max}$. Thus the only admissible solution
of Eq.(\ref{eq22}) is $(u=u_{max},\, v=0)$, i.e. the positions of
the point-like charges. The rest of the lobes is cut because the
mapping domain is restricted to have a single-valued potential
matching the Coulombian potential at the infinity.

In order to compute the force (\ref{eq17}) we will take into
account that the points on the $y$-axis satisfy
\begin{equation}
\label{eq23} \frac{u}{2\lambda }\,\;<\;\,0\,\,\,\,\;\quad
\,\mbox{and}\,\,\quad \,\,\,\frac{v}{2\lambda }=\pm \pi .
\end{equation}
Then, according to Eq.(\ref{eq21}), the positive $y$-semiaxis can
be parametrized by defining a parameter $t$ such that $\cos
(t/2)=\exp [u/(4\lambda )]$; thus it results
\begin{equation}
\label{eq24} y(t)=-\frac{d}{2}\;\,\left[ {\tan \, {\frac{t}{2}}
\;-\;\frac{4}{\alpha ^2}\;\,(\pi +t-\sin \;t)} \right]
\end{equation}
where $t \in (-\pi, t_{o})$. Then the integral (\ref{eq17}) can be
computed as twice the integral along the positive $y$ semi-axis.
Thus the force is
\begin{eqnarray}
\nonumber F^x=\frac{2\lambda }{\pi \,d}\int\limits_{-\infty
}^{\;\;\quad u_o } {\;\exp \left[ {\frac{u}{2\lambda }}
\right]\;\;\sqrt {\exp \left[ {-\frac{u}{2\lambda }} \right]-1}
\;\,du\,} \\ =\,\;\frac{4\lambda ^2}{\pi \,d}\int\limits_{-\pi
}^{\,\,\,\,\,\,\,t_0 } {\sin ^2\, {\frac{t}{2}}\,}
dt\,=\;\,\,\frac{2\,\lambda ^2}{\pi \,d}\;\left( {\pi +t_o -\sin
\,t_o } \right)\,\label{eq25}
\end{eqnarray}
In the former expression, $u_{o}$ is such that $y(u_{o}) = 0$. In
the Coulombian case ($b \to \infty $, i.e. $\alpha \to \infty$) it
is $u_{o} = 0$, so $t_{o} = 0$ and the Coulomb force is recovered.
Instead, in Born-Infeld theory it is $u_{o} < 0$ and $-\pi < t_{o}
< 0$. The dependence of the interaction on $b$ is given through
the value $t_{o}$ which depends on $\alpha = b d/\lambda$. Since
$t_{o }- \sin t_{o}$ is negative for $-\pi <  t_{o} < 0$, it is
concluded that the interaction between equal monopoles is
repulsive for all value of $\alpha $, but less intense than the
Coulombian interaction. In particular, for $\alpha \to 0$, it is
$t_{o} \to -\pi$; thus the repulsive interaction force vanishes
for $d\to 0$.

In order to get an explicit correction to the Coulombian force we
will take into account that $t_{o}$ is small for high values of
$\alpha $. Thus we can try to solve $y(t_{o}) = 0$ in (\ref{eq24})
by writing $t_{o}$ as a power series of $\alpha^{-2}$. In this way
we reach the result $t_{o }=-8\, \pi\, \alpha^{-2}+ 128\, \pi^3\,
\alpha^{-6 }/3- 2048\, \pi^3\, \alpha^{-8}/3 + O(\alpha^{-10})$.
Therefore the force (\ref{eq25}) is
\begin{equation}
\label{eq26}
F^x\;\;\mathrel{\mathop{\kern0pt\longrightarrow}\limits_{\alpha
\to \infty }} \;\,\frac{2\lambda ^2}{d}\;\left(
{\,1-\;\,\frac{256\;\pi ^2}{3\;\alpha ^6}} \right)\, +\,
O\mbox{(}\alpha ^{-10}\mbox{)}.
\end{equation}
Then, Born-Infeld correction to the repulsive force between equal
charges is very weak. Notice that $d$ in (\ref{eq26}) is not the
real distance $D$ between the charges. $D$ goes to $d$ when $b \to
\infty$, but $D$ is smaller than $d$ for equal charges (see Figure
\ref{Fig3}):
\begin{equation}
D_{repulsive}\,=\,d\,-\,\frac{2\,d}{\alpha^2}\, \left(\ln
(\alpha^2)-1\right)\,+\,O[\alpha^{-4}]\ .
\end{equation}

\subsection{Opposite charges}
\begin{figure}[ht]
\centering
\includegraphics[scale=.3]{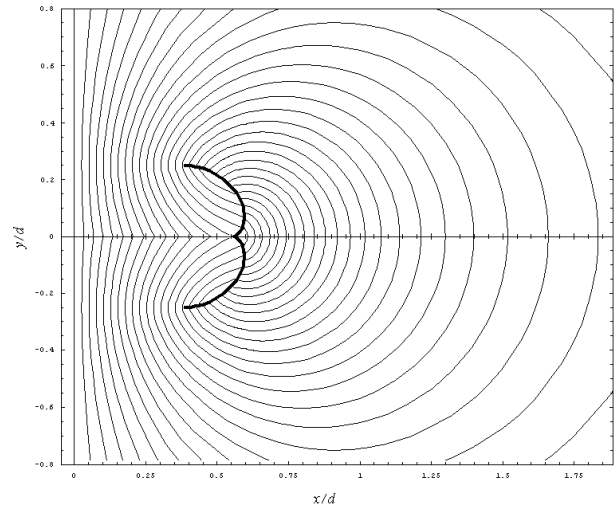}
\caption[]{Born-Infeld equipotencial lines ($u=constant$) for
$\alpha=5$, together with the curve describing the right
charge.}\label{Fig4}
\end{figure}

\begin{figure}[ht]
\centering
\includegraphics[scale=.4]{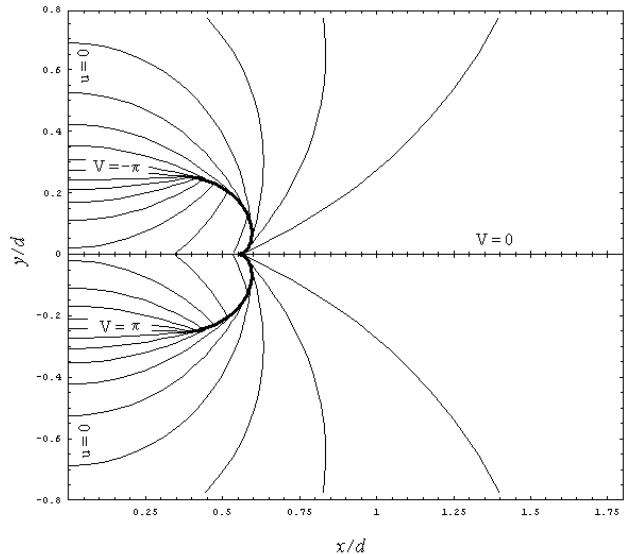}
\caption[]{Born-Infeld field lines ($v=constant$) for $\alpha=5$,
together with the curve describing the right charge. In the figure
$V$ stands for $v/(2\, \lambda)$.}\label{Fig5}
\end{figure}

We will now repeat the former steps for the case of the attractive
charge configuration consisting of a charge $-\lambda $ at $x =
-d/2$, and a charge $\lambda $ at $x = d/2$. The complex
Coulombian potential is
\begin{equation}
\label{eq27} w = -2\lambda\, \ln \left(
{\frac{\frac{2z}{d}-1}{\frac{2z}{d}+1}} \right).
\end{equation}
Here we have chosen the integration constant such that $w(z)$ is
null at infinity. Inverting (\ref{eq27}) we obtain the Coulombian
mapping
\begin{equation}
\label{eq28} z\ =\ \frac{d}{2}\,\coth \,{\frac{w}{4\lambda }}\ ,
\end{equation}
This means that function $p(w)$ is
\begin{equation}
\label{ee4} p(w) = -\frac{d}{8\,\lambda}\,
\sinh^{-2}\frac{w}{4\,\lambda}.
\end{equation}
Thus the Born-Infeld mapping results
\begin{equation}
\label{eq29} z=\frac{d}{2}\left[ {\coth \, {\frac{w}{4\lambda }}
+\frac{4}{\alpha ^2}\left( {\frac{\overline w}{2\lambda }-\sinh \,
{\frac{\overline w}{2\lambda }}} \right)} \right].
\end{equation}
The points where the electrostatic field reaches the value $b$
(i.e., $|p(w)|=(2\, b)^{-1}$) belong to the curve
\begin{equation}
\label{eq30} \frac{u}{2\lambda }=\pm\; {\rm arccosh}\,\left(
{\frac{\alpha}{2}+\cos \, {\frac{v}{2\lambda }}} \right).
\end{equation}
If $\alpha >4$ this curve decomposes in two separate parts
(``charges'') at each side of the $y$-axis. Like the previous
case, the Born-Infeld mapping (\ref{eq29}) fails to be periodic
due to the presence of a linear term; so the $u-v$ domain in
mapping (\ref{eq29}) should be properly restricted to get a
single-valued potential $u(x, y)$. The branch of the complex
potential $w$ to be kept is those matching the Coulombian
potential at infinity. Differing from the previous case, this
branch does reach the curves (\ref{eq30}). Figures \ref{Fig4} and
\ref{Fig5} show the equipotential ($u=constant$) and field
($v=constant$) lines surrounding the right charge, and their
relation with curve (\ref{eq30}). Again the lines $v=0$ coincide
with the piece of the $x$-axis going between the charges and
infinity. But in this case the domain of $|v|/(2\,\lambda)$ has to
be extended beyond $\pi$ to reach the piece of the $x$-axis
joining the charges. Both the potential $u$ and the field $\bf E$
are discontinuous at the charge (the field attains the maximum
value $b$ at the exterior side of the charge, i.e. the side where
$0\le |v|/(2\,\lambda)\le\pi$). Besides, the field is
discontinuous at the branch cut on the $x$-axis between the
charges.

\begin{figure}[ht]
\centering
\includegraphics[scale=.8]{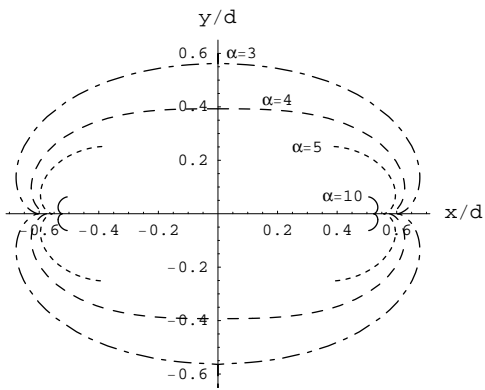}
\caption[]{Born-Infeld opposite ``charges'' for $\alpha=10$ and
$\alpha=5$. For $\alpha\le 4$ the charges merge in a sole
object.}\label{Fig6}
\end{figure}

For $\alpha \le 4$ the curve of maximum field get closed, as it is
typical for a dipole \cite{21}. In this case, the ``charge''
distribution becomes a unique object; the Coulombian zone $|p(w)|>
(2b)^{-1}$ is, of course, the outside of the object. Figure
\ref{Fig6} shows the curves $|p(w)|= (2b)^{-1}$ for different
values of $\alpha$. The curves display cusps on the $x$-axis
($v=0$) at $\vert x\vert
> d/2$. At the cups it is $u=\pm 2 \lambda\; {\rm arccosh}\,(
\alpha/2+1)$.

We will calculate the force (\ref{eq18}) for $\alpha>4$, which is
the case where the charges are separate. Since the $y$-axis is
characterized by $u=0$, then Eq.(\ref{eq29}) says that the
positive $y$-axis accepts a parametrization similar to
(\ref{eq24}) whenever the parameter $t$ is defined as
$t=v/(2\lambda)-\pi$:
\begin{equation}
y(t)\ =\ -\frac{d}{2}\,
\left[\tan\frac{t}{2}-\frac{4}{\alpha^2}(\pi+t+\sin
t)\right]\label{eq31}
\end{equation}
This function is monotonous for $t \in (-\pi, t_{o})$, where $0 <
t_{o} < \pi$ is the parameter satisfying $y(t_{o}) = 0$. Since
$p(w)\vert _{y-axis} =d/(8\lambda )\,\,\sin ^{-2}\left[
{v/(4\lambda )} \right]$, then the force (\ref{eq18}) is
\begin{equation}
\label{eq32} F^x = -\frac{4\,\lambda ^2}{\pi \,d}\int\limits_{-\pi
}^{t_o} {\cos ^{\,2}\frac{t}{2}\,dt} = -\,\frac{2\,\lambda ^2}{\pi
\,d}\,(\pi + t_o + \sin\,t_o)
\end{equation}
For $b \to \infty $ ($\alpha \to \infty $), it is $t_{o }\to  0$
in Eq.(\ref{eq32}), so the Coulombian force is recovered. Since
$t_{o} \in  (0, \pi)$, it is concluded that the attraction between
opposite monopoles is more intense than Coulombian interaction. We
will solve $y(t_{o}) = 0$ in (\ref{eq31}) by writing $t_{o}$ as a
power series of $(\alpha^2-16)^{-1}$. The result is $t_{o }=
8\pi\, (\alpha^2-16)^{-1}\,(1-16\pi^2(\alpha^2-16)^{-2}/3) +
O[(\alpha^2-16)^{-4}]$. Therefore the Born-Infeld interaction
(\ref{eq32}) between opposite monopoles behaves as
\begin{eqnarray}
\label{eq33} \nonumber
F^x\;\;\mathrel{\mathop{\kern0pt\longrightarrow}\limits_{\alpha
\to \infty }}\,\frac{2\lambda ^2}{d}\,\left(
{1\,+\,\frac{16}{\alpha ^2-16}-\,\frac{512\,\pi ^2}{3(\alpha
^2-16)^3}} \right)\\ +\, O\left[(\alpha ^2-16)^{-4}\right].
\end{eqnarray}
Differing from the repulsive case, the attractive interaction
receives a more perceptible correction of order
$(\alpha^2-16)^{-1}$. Notice that $d$ in (\ref{eq33}) is not the
distance $D$ between the cusps in Figure \ref{Fig6}. By computing
the positions of the cusps for opposite charges, it results that
$D$ is larger than $d$:
\begin{eqnarray}
\nonumber D_{attractive }\,=\,d\, +\, \frac{2\, d}{\alpha^2-16}\,
\left(\ln (\alpha^2-16)\, -\, 3\right)\\ +\,
O[(\alpha^2-16)^{-3/2}]\ .
\end{eqnarray}

\section{Conclusions}
The first goal of Born-Infeld theory was the obtaining of a
point-like charge solution with finite self-energy. However this
solution has intriguing features: $\partial L_{BI}/\partial {\bf
E}$ still diverges, and $\bf E$ is finite at the charge position
(an unpleasant property for a vector field at its center of
symmetry). However, these disagreeable features do not cause any
trouble to the interaction between charges. We have considered a
``two charges'' field (which differs from the mere superposition
of two ``one charge'' fields, since the theory is nonlinear). To
compute the interaction between parts of this field configuration
one must consider the momentum flux through a surface separating
both subsystems. According to the method developed in Ref.
\cite{21}, the interaction force is given by expressions (17-18),
where $p$ is a Coulombian function, and the Born-Infeld features
are encoded in the integration interval. Although we have obtained
the force in a parametric form (the parameter $t_{o}$ in forces
(\ref{eq25}) and (\ref{eq32}) comes from the transcendent equation
$y(t_{o}) = 0$ in (\ref{eq24}) and (\ref{eq31}) respectively), we
have succeeded in computing Born-Infeld corrections to Coulombian
interactions. In addition we have proved that the interaction
force between equal charges is well behaved and goes to zero when
the charges approach. This limit cannot be reached for opposite
charges because they merge in a unique dipolar object of finite
size.

\acknowledgments{R.F. wishes to thank G. Giribet, J. Oliva and R.
Troncoso for helpful comments. This work was partially supported
by Universidad de Buenos Aires (Proy. UBACYT X103) and CONICET
(PIP 6332).}

\end{document}